\documentclass[apj]{emulateapj}\usepackage{apjfonts}\newcommand{\escl}[1]{}\newcommand{\epsstr}[1][1.15]{\epsscale{#1}}
\usepackage{natbib, graphicx}
\bibpunct[, ]{(}{)}{;}{a}{}{,}
\newcommand{\mmbox}[1]{\mbox{$#1$}}
\newcommand{\dif}{\mathrm{d}}
\newcommand{\diff}[2]{\frac{\dif#1}{\dif#2}}
\newcommand{\pma}[2]{{}^{+#1}_{-#2}}
\newcommand{\xil}{\xi_\mathrm{L}}
\newcommand{\nxi}{\hat{\xi}}
\newcommand{\xib}{\xi_\mathrm{bin}}
\newcommand{\xis}{\xi_\mathrm{sys}}
\newcommand{\pct}{\%}
\newcommand{\unit}[1]{\mbox{$\,\mathrm{#1}$}}
\newcommand{\abd}{\alpha_\mathrm{BD}}
\newcommand{\rbod}{\mathcal{R}}
\newcommand{\rpop}{\mathcal{R}_\mathrm{pop}}
\newcommand{\rkbd}{\mathcal{R}_\mathrm{HBL}}
\newcommand{\nbod}{N_\mathrm{bod}}
\newcommand{\nsng}{N_\mathrm{sng}}
\newcommand{\nsys}{N_\mathrm{sys}}
\newcommand{\nsbd}{N_\mathrm{sys,BD}}
\newcommand{\nsmd}{N_\mathrm{sys,MD}}
\newcommand{\nsst}{N_\mathrm{sys,star}}
\newcommand{\nbny}{N_\mathrm{bin, tot}}
\newcommand{\nbbd}{N_\mathrm{BD-BD}}
\newcommand{\nbmd}{N_\mathrm{MD-MD}}
\newcommand{\nbst}{N_\mathrm{star-star}}
\newcommand{\nbsb}{N_\mathrm{star-BD}}
\newcommand{\mmin}{m_\mathrm{min}}
\newcommand{\mmax}{m_\mathrm{max}}
\newcommand{\msys}{m_\mathrm{sys}}
\newcommand{\mbin}{m_\mathrm{bin}}
\newcommand{\mpri}{m_\mathrm{prim}}
\newcommand{\ma}{m_\mathrm{A}}
\newcommand{\mb}{m_\mathrm{B}}
\newcommand{\msun}{M_{\sun}}
\newcommand{\mh}{m_\mathrm{H}}
\newcommand{\mlobd}{m_\mathrm{0,BD}}
\newcommand{\mlost}{m_\mathrm{0,star}}
\newcommand{\mhibd}{m_\mathrm{max,BD}}
\newcommand{\ftot}{f_\mathrm{tot}}
\newcommand{\fbd}{f_\mathrm{BD-BD}}
\newcommand{\fst}{f_\mathrm{star-star}}
\newcommand{\fmd}{f_\mathrm{MD-MD}}
\newcommand{\fmix}{f_\mathrm{star-BD}}
\newcommand{\tmsun}{\mbox{$\msun$}}
\newcommand{\simf}{\mbox{$\mbox{IMF}_\mathrm{sys}$}}
\newcommand{\bmf}{\mbox{$\mbox{IMF}_\mathrm{bin}$}}

\journalinfo{The Astrophysical Journal}
\submitted{Accepted 2007 August 10}
\begin{document}
\title{A Discontinuity in the Low-Mass Initial Mass Function}
\shorttitle{A discontinuity in the low-mass IMF}
\author{Ingo Thies\altaffilmark{1} \& Pavel Kroupa\altaffilmark{1}}
\shortauthors{Thies \& Kroupa}
\altaffiltext{1}{Argelander-Institut f\"ur Astronomie (Sternwarte), Universit\"at Bonn, Auf dem H\"ugel 71, D-53121 Bonn, Germany}
\begin{abstract}
The origin of brown dwarfs (BDs) is still an unsolved mystery.
While the standard model describes the formation of BDs and stars in a
similar way recent data on the multiplicity properties of stars and
BDs show them to have different binary distribution functions.
Here we show that proper treatment of these uncovers a discontinuity
of the multiplicity-corrected mass distribution in the very-low-mass star
(VLMS) and BD mass regime.
A continuous IMF can be discarded with extremely high confidence.
This suggests that VLMSs and BDs on the one hand, and stars on the other,
are two correlated but disjoint populations with different dynamical
histories.
The analysis presented here suggests that about one BD forms per five stars
and that the BD-star binary fraction is about 2\pct--3\pct\ among stellar
systems.
\end{abstract}
\maketitle
\section{Introduction}
\label{sec:intro}
Traditionally, brown dwarfs (BDs) are defined as (sub)stellar bodies with
masses below the hydrogen burning mass limit (HBL),
$\mh=0.075\,\msun$ for solar composition, and consequently they cool
indefinitely after formation \citep{Buetal93,ChaBa00}.
Several attempts have been made to
explain the formation of BDs by the same mechanisms as for
stars, i.e. via fragmentation of a gas cloud and subsequent
accretion \citep{AdFa96,PaNo02,PaNo04}.
If a gas cloud in a star-forming region fragments there will be a
certain number of gas clumps with masses below the HBL.
Unless the mass of the fragment is below the local Jeans mass
it will contract in essentially the same way as higher mass clumps and
finally produce a single or multiple BD.
This scenario predicts similar multiplicities and also a substellar
initial mass function (IMF) as a continuous extension of the stellar one
(e.g. the standard model with BDs in \citealt{Ketal03}).

However, recent observations have shown that there
is a lack of BD companions to low-mass stars
\citep*{2003IAUS..211..279M}. \citet{GreLin06} found a star-BD binary
fraction among solar-type primaries of less than 1\pct\ for close
companions,
the \emph{brown dwarf desert}. This implies two populations of stellar
and substellar objects, and that binaries are formed in
each population separately (except for pairing due to post-formation
dynamical exchanges). Observations e.g. by \citet{Reidetal06}
also show that most BD binaries have a primary-to-companion mass ratio
of $q>0.8$, in contrast to the mass ratio distribution of stellar binaries
which has typically $q<0.4$ \citep{DuqMay91}.

There are indications, e.g. \citet{MetHil05}, that the
BD desert may not be as dry for larger separations ($>30$~AU) as it is
for smaller ones. Since new surveys using adaptive optics or new instruments
like the upcoming James Webb Space Telescope might reveal more substellar
companions to stars, the fraction of star-BD systems may increase.

Apart from the BD desert there are more hints for a separate population.
For example, BDs and VLMSs have a relatively low binary fraction
of about 15\pct\
(\citealt{Bouyetal03,Cloetal03,Maetal03}; \citealt*{Krausetal06};
\citealt{LaHoMa07}).
By comparison, the stellar binary fraction
is close to 100\pct\ for the very young Taurus-Auriga association (TA, about
1~Myr; \citealt{Duchene99,Luetal03a}) and about 40\pct--50\pct\ for other clusters
and field stars \citep{Lada2006}.
The BD and VLMS binary fraction can be increased to a starlike binary
fraction if there are a large fraction of $\la5$~AU binaries,
e.g. as deduced by \citet{JefMax05}. But such a semi-major axis distribution
would again imply a discontinuity of its form between low-mass stars and
VLMSs/BDs and is not supported by the radial-velocity survey of
\citet{Joergens2006a}.
We therefore do not consider the starlike formation as a major mechanism
for BDs.

It has also been argued that
the low binary fraction of BDs can be understood as a continuous extension
of a trend that can already be recognized from G dwarfs to M dwarfs
\citep{Lu04b,SteDur03}.
Therefore, the binary fraction alone cannot be taken as a strong evidence
to introduce a separate population.

\begin{figure}
\begin{center}
\epsstr
\plotone{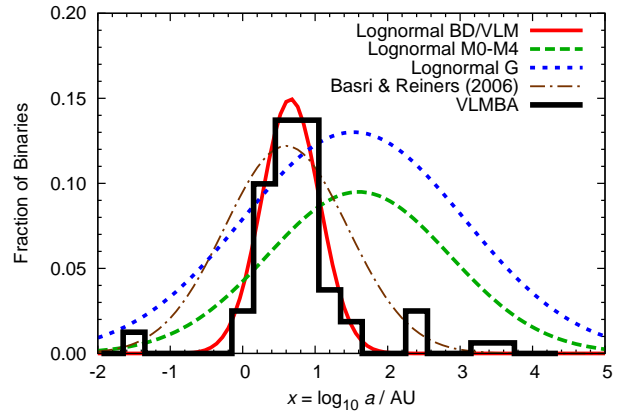}
\caption{\label{sepdist}The semi-major axis ($a$) distribution for
VLM binaries (VLMBs) with a total mass $\mbin<0.2\,\msun$
based on data from Nick Siegler's
\emph{Very Low Mass Binaries Archive}
(VLMBA, http://paperclip.as.arizona.edu/\symbol{126}nsiegler/VLM\_binaries/,
version from June 1, 2007; histogram).
The data can be fitted with a $\log$-normal distribution.
For the VLMBs (solid curve) the
Gaussian parameters are $x_\mathrm{peak}=0.66$ (corresponding to
$a=4.6\unit{AU}$) and $\sigma=0.4$ with a
normalization factor (= total binary fraction) $\ftot=0.15$,
the binary fraction among BDs and VLMSs. For the M0--M4 dwarf
binaries by \citet{Cloetal03} $x_\mathrm{peak}=1.6$ and $\sigma=1.26$
($\ftot=0.3$) and for the
G dwarf binaries $x_\mathrm{peak}=1.53$ and $\sigma=1.53$ ($\ftot=0.5$;
\citealt{FisMar92}, long-dashed and short-dashed curves, respectively).
For VLMBs the gap within $-1.5\le\log a/AU\le0$
can be interpolated with the compressed Gaussian of \citet{BasRei2006}
with $\sigma=0.85$ (thin dot-dashed curve),
also enveloping wider BD/VLMS binaries found by \citet{Konetal07}.
It corresponds to a total BD/VLMS $\fbd=0.26$.
The areas below the histogram and the curves are equal to
the corresponding $\ftot$ (eq. \ref{ftot}).}
\end{center}
\end{figure}

\begin{figure}
\begin{center}
\epsstr
\plotone{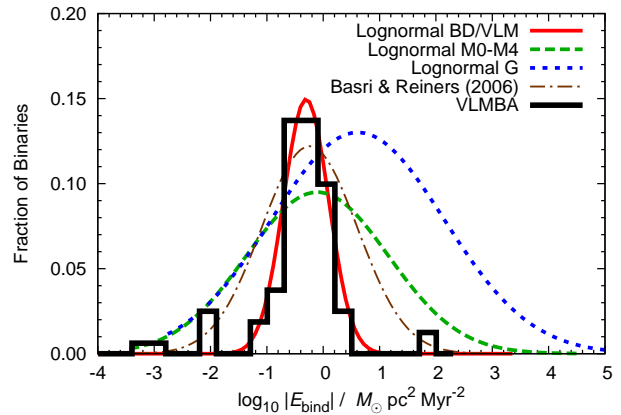}
\caption{\label{edist}The distribution of the orbital binding
energy $E_\mathrm{bind}=-0.5 G(m_1 m_2/a)$ for VLM binaries and low-mass
stellar binaries. Line types and symbols are the same as in Fig.
\ref{sepdist}.
As for the semi-major axes there is a clearly
different distribution for BDs/VLMSs on the one hand side
and M, K and G dwarf binaries on the other.}
\end{center}
\end{figure}

The most striking evidence for two separate populations is the empirical fact
that the distribution of the separations and therefore the binding energies
of BD binaries differs significantly from that in the stellar regime
\citep{Bouyetal03,Burgetal03,Maetal03,Cloetal03}.
This is shown in Figures \ref{sepdist} and \ref{edist}.
The solid line is a Gaussian fit to the central
peak of the histogram while the dash-dotted one refers to
(\citealt{BasRei2006}, a compressed \citealt{FisMar92} fit).
BDs and very low-mass stars (VLMS) have a semi-major axis distribution limited
to \mmbox{\lesssim15}~AU, whereas M, K, and G dwarfs have a very
broad and similar distribution (\emph{long-dashed and short-dashed curves};
\citealt{Cloetal03}; \citealt*{LaHoMa07}; \citealt{PPVGoodwinetal}).
There is also a dearth of BDs below 1~AU. Recent findings, e.g. by
\citet{GuWu03} and \citet{Kenetal05}, suggest a low number of such very close
BD/VLMSs binaries. The semi-major axis distribution
of BDs/VLMSs binaries based on the data from \citet{Cloetal03} can be
modelled with a log~$a$ Gaussian centered at 4.6~AU ($\log a =0.66$) with
a half-peak width of $\sigma=0.4$. It corresponds to an overall BD/VLMS
binary fraction of $\fbd=0.15$. If data from \citet{Lu04a},
\citet{Joergens2006a}, and \citet{Konetal07} are taken as hints to
incomplete data between about 0.02 and 1~AU,
the compressed \citet{FisMar92} Gaussian from Figure 4 in
\citet{BasRei2006} may provide an appropriate envelope. However, for an assumed
BD mass of 0.07~\tmsun\ and $\fbd=0.26$ their period distribution corresponds
to a semi-major axis distribution with $\sigma\approx0.85$ and is therefore
still inconsistent with that of M and G dwarfs.

Although \citet{Konetal07} have recently found five VLMS binaries in TA
with four of them having separations much larger than 15~AU,
the sudden change of the orbital properties
remains. In particular, they found two binaries with projected
separations slightly above 30~AU and two others with separations
between 80 and 90~AU. Possible implications of these discoveries
are discussed in \S~\ref{ssec:diskfrag}.
However, the truncation near 15~AU cannot be derived from the stellar
distribution through downsizing according to Newton's laws
\citep{Cloetal03,Bouyetal03}.
Not even dynamical encounters in dense stellar environments can invoke such
a truncation near 15~AU \citep{Burgetal03}.
Using $N$-body simulations \citet{Ketal03} tested the hypothesis that BDs
and stars form alike, and mix in pairs, and found that, despite of close
dynamical encounters,
the distribution of the semimajor axes of BD binaries remains starlike.
That is, dynamical encounters even in dense clusters cannot truncate
the BD binary distribution near 15~AU.
They further found that star-BD binaries would be much more frequent
than actually observed. Thus, this classical hypothesis is rejected with
high confidence.

With this contribution we study the implications of the observed change
of binary properties on the underlying single-object initial mass function
(IMF). For this purpose, we analyzed the observational mass functions of TA
(Luhman et al. 2003a, 2004b), the IC~348 cluster
\citep{Luetal03b},
the Trapezium cluster \citep{Muetal02}, and the Pleiades cluster based on
data by \citet{Doetal02}, \cite{Moetal03}, and the Prosser and Stauffer
Open Cluster
Database.\footnote{Available at
http://www.cfa.harvard.edu/\symbol{126}stauffer/opencl/}
For all the systems we analyst, we refer to the MF as the IMF, although
strictly speaking this is not correct for the 130~Myr old Pleiades.
In \S~\ref{sec:IMF} we shortly review the definition of the IMF
and the role of multiplicity on the shape of the
individual body IMF compared to the observed system IMF (\simf).
We show that the IMF that can be derived from an observed
\simf\ does not need to be continuous even if the \simf\ does not show any
discontinuity.
In \S~\ref{sec:method} we describe the statistical method how to
fit a model \simf\ by combining BD and star IMFs
as an approximation to the observed \simf.
\S~\ref{sec:results} presents the results that indicate a
discontinuity close to the HBL. Also the BD to star ratio from the
model is calculated there.
In \S~\ref{sec:discussion} the results are discussed in the context
of four alternative BD formation scenarios, i.e. embryo ejection,
disk fragmentation, photoevaporation, and ejection by close stellar
encounters.

\section{The IMF for individual stars and systems}
\label{sec:IMF}
\subsection{Definition}
\label{ssec:IMF:def}
The IMF is among the most important properties of
a stellar population since it gives hints to the processes that form
stars and BDs. Although we can only observe stellar populations at their given
age there are data of several very young populations where the mass function
is probably still very close to the initial one.

In general, the mass distribution of stars and BDs can be approximated by
a power law or a combination of several power laws.
\citet{Salpeter1955} estimated the relationship between the stellar
mass and the relative number of stars of a given mass as a single power law,
\begin{equation}\label{powerimf}
\xi(m)=\diff{n}{m}=k\,m^{-\alpha}\,,
\end{equation}
for $0.4\lesssim m/\msun\lesssim 10$ and with $\alpha=2.35$ and a normalization
constant $k$.
The IMF is often expressed on the logarithmic mass scale and
then becomes
\begin{equation}\label{powerlog}
\xil(\log m)=\diff{n}{\log m}=\left(\ln 10\right)\,m\,\xi(m)=k_\mathrm{L}\,m^{1-\alpha}\,,
\end{equation}
where $k_\mathrm{L}$ is the corresponding normalization constant for $\xil$.
An IMF is said to be \emph{flat} if $\xil(\log m)$ is constant, i.e.
$\alpha=1$.
More recent work has shown there to be a flattening in the lower-mass regime
of the observed mass function and, in the $\xil$ representation,
even a turnover near the BD-star transition \citep*{Kr01,RGH02,Cha03}.
Here, all objects from BDs to the most massive stars
in a cluster are described by a \emph{continuous IMF}, i.e.
with a single population containing BDs as well as stars.
The universal or ``canonical'' IMF has $\abd\equiv\alpha_0=0.3$
($0.01\le m/\msun\le0.075$), $\alpha_1=1.3$ ($0.075\le m/\msun\le0.5$), and
$\alpha_2=2.3$ ($0.5\le m/\msun\le\mmax$), where $\mmax$ is given by the
mass of the host cluster \citep{WK06}. The generally accepted wisdom has been
that the IMF is continuous from above 0.01~\tmsun\ to $\mmax$
(e.g. \citealt{Cha02}).

\subsection{Unresolved Binaries}
\label{ssec:IMF:bincorr}
Star cluster surveys are usually performed with wide-field telescopes with
limited resolution that do not resolve most of the binaries. Hence, they yield,
as an approximation, system MFs
in which unresolved binary systems are counted as one object.
The fraction of
unresolved multiples can be taken as the total binary fraction of a cluster,
\begin{equation}\label{ftot}
\ftot=\frac{\nbny}{\nsng+\nbny}\,,
\end{equation}
where $\nsng$ is the number of singles (or resolved individual bodies)
and $\nbny$ the number of (unresolved) binaries.
Unresolved binaries increase the number of individual bodies, $\nbod$, in a
cluster such that
\begin{equation}\label{defnbod}
\nbod=\left(1+\ftot\right)\nsys\,,
\end{equation}
where $\nsys=\nsng+\nbny$ is the total number of systems.
Note that a ``system''
is either a single body or a binary or a higher order multiple. Here we
ignore higher order multiples, because they are rare \citep{GoKr05}.
The binary fraction in dependence of the mass of the
primary star, $\mpri$, is
\begin{equation}\label{fmpri}
f(\mpri)=\frac{\nbny(\mpri)}{\nsng(\mpri)+\nbny(\mpri)}\ ,
\end{equation}
where $\nbny(\mpri)$ and $\nsng(\mpri)$ are, respectively, the number of
binaries with primary star mass $\mpri$ and single stars of mass $\mpri$.

\subsection{The System Mass Function}
\label{ssec:IMF:SIMF}
The effect of this \emph{binary error} on the appearance of the IMF can
be described as follows.
Assume a stellar population with $\ftot<1$ and
with stellar masses with a minimum
mass $\mmin$ and a maximum mass limit of $\mmax$. The minimum
mass of a binary is $2\mmin$, while the mass-dependent binary fraction
$f(m)=0$ for $\mmin\le\msys<2\mmin$.
A binary closely above $2\mmin$ can only consist of two stars near
$\mmin$ making such binaries rare.
For higher system masses, where a system can be a multiple or
a single star, there are more possible combinations
of primary and companion mass, so that the binary fraction
increases with the system mass and approaches
an upper limit for the most massive objects.

\begin{figure}
\begin{center}
\epsstr
\plotone{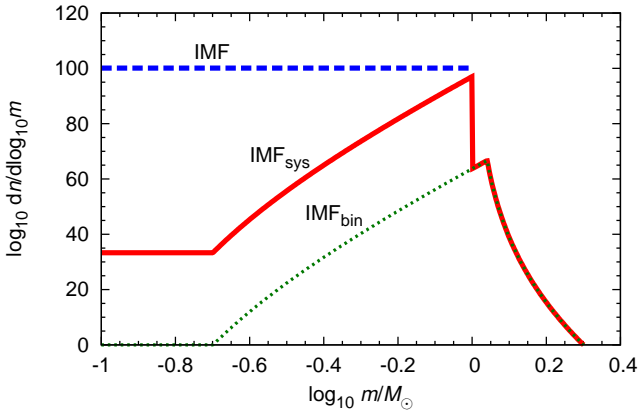}
\caption{\label{bincorr}The system IMF (\simf\, solid line) corresponding
to a flat logarithmic IMF (i.e. $\alpha=1$, dashed line) for
$\mmin=0.1\le m/\msun\le 1=\mmax$ with an overall binary fraction of 50\pct.
The corresponding binary IMF (\bmf\, eq. \ref{defbmf}) is shown by the thin
dotted line. The system MF peaks at 1~\tmsun\ and is truncated for higher
masses while there is a minor peak at 1.1~\tmsun\ corresponding to the
peak of the binary MF at \mmbox{\mmin+\mmax}.}
\end{center}
\end{figure}

Figure \ref{bincorr} shows the general shape of a system IMF for a
flat (logarithmic scale) IMF with $\mmin=0.1\,\msun$, $\mmax=1\,\msun$, and
$\ftot=0.5$. The \simf\ is flat between $\mmin$ and $2\mmin$
($\log m/\msun\approx-0.7$) and rises above $2\mmin$ to a maximum
at $\mmax$.
Systems with $\msys<2\mmin$ can only be singles and the \simf\ in this
region is just the IMF minus the mass function of objects that are bound
to a multiple system.
For masses $\msys>\mmax$, on the other hand, only binaries exist,
and the \simf\ declines towards zero at $\msys=2\mmax$, the highest
mass possible for binaries. The sharp truncation of the
IMF at $m=\mmax$ causes the sudden drop at $\msys=\mmax$,
while the minor peak at $\msys=\mmax+\mmin$ corresponds to the maximum of
the \emph{binary mass function} (\bmf, the IMF of binary system masses).
For a binary of this mass the primary and companion mass can be drawn from
the whole IMF, and thus the number of possible combinations becomes maximal.
It should be noted that natural distributions
with smoother boundaries probably do not show such a double peak.

Mathematically and in the case of random pairing, which is a reasonable
approximation in the stellar regime \citep*{MalZin01,PPVGoodwinetal},
the binary mass function
is just
the integral of the product of the normalized IMF, $\nxi\equiv\xi/\nbod$, of
each component times the total number of binaries $\nbny$.
Given
the masses of the binary components A and B, $\ma$ and $\mb$, the binary
mass \mmbox{\mbin=\ma+\mb}. Thus, \mmbox{\mb=\mbin-\ma}. Thus, \bmf\ can
now be written as
\begin{equation}\label{defbmf}
\xib(\mbin)=\nbny\int\limits_{\mmin}^{\mbin-\mmin} \nxi(m)\,\nxi(\mbin-m)\,\dif m\,,
\end{equation}
where $\mmin$ is the lower mass limit of all individual bodies in the
population. The upper limit of the integral, $\mbin-\mmin$, is the maximum
mass of the primary component $\ma$ corresponding to a secondary component
with $\mb=\mmin$.

The other extreme case of assigning the component masses is
equal-mass pairing. In that case, equation (\ref{defbmf}) simplifies to the
IMF of one of the components and \bmf\ is just the
IMF shifted by a factor of 2 in mass and corrected for binarity
using equations (\ref{ftot}) and (\ref{defnbod}):
\begin{equation}
{\xib}_\mathrm{equal}(\mbin=2m)=\frac{\ftot}{1+\ftot}\,\xi(m)\,.
\end{equation}
This case is more applicable for BD binaries since their mass ratio
distribution peaks at a ratio $q=1$ \citep{Reidetal06}.
However, due to the low overall binary
fraction of BDs the effect of the mass ratio distribution is quite small.

The \simf\ is just the sum of \bmf\ and the IMF of the remaining
single objects, $(1-\ftot)\xi$:
\begin{equation}\label{cimf}
\xis(m)=\xib(m)+\left(1-\ftot\right)\,\xi(m)\,.
\end{equation}

Thus, to obtain the true individual star IMF, $\xi(m)$ has to be extracted
from the observed $\xis(m)$ for which a model for $\xib(m)$ is required.
That is \simf\ has to be corrected
for unresolved binaries in the cluster or population under study.
This leads to a significant increase of the numbers of objects at the
low-mass end because low-mass objects contribute to both low-mass singles
and intermediate-mass binaries and thus more individual objects are
required to reproduce the observed \simf\ \citep{KGT91,MalZin01}.
In the mass range $\mmin$--$2\mmin$,
systems can only be single because the system mass is the sum
of the masses of the system members. Increasing the system mass
beyond $2\mmin$ causes the binary contribution to
rise quickly and then to asymptotically approximate a maximum value.
Thus, the fraction of singles among M dwarfs
is higher than for G dwarfs \citep{Lada2006},
as one would expect for a stellar population ($>0.08\,\msun$), i.e.
essentially without BDs.

\begin{figure}
\begin{center}
\epsstr
\plotone{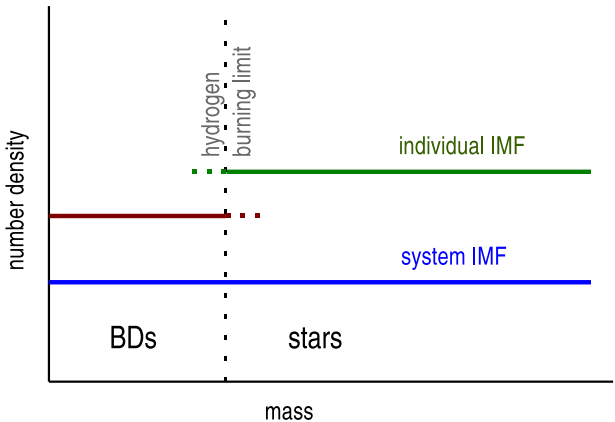}
\caption{\label{cartoon}A population of stars and BDs
with different binary fractions can result in a discontinuous IMF,
even if the observed (system) IMF appears to be continuous.
The binary fraction of the BDs is lower than that of stars and therefore
a lower number, $\nbod$, of individual objects is required for the frequency
of BD and stellar systems of a given mass being equal (eq. \ref{defnbod}).
Note the overlap region
indicated by the horizontal dotted lines: some starlike bodies may actually
be physical BDs (upper line) while some BD-likes are indeed very-low-mass stars.
}
\end{center}
\end{figure}

Figure \ref{cartoon} schematically shows the effects of correcting
a flat continuous observed \simf\ for unresolved multiples whereby the
binary fraction for BDs is smaller than for stars, as is observed to be
the case.
For a binary fraction
of 15\pct\ (i.e. $\fbd=0.15$) among BDs there would thus be
\mmbox{N_\mathrm{BD}=(1+\fbd)\nsbd=1.15\nsbd} BDs in total, while
for stars with \mmbox{\ftot=0.5} we would have \mmbox{1.5\nsst}.
This exemplifies how the change  in $f$ leads to a discontinuous \simf.

\section{The Method}\label{sec:method}
\begin{table}
\caption{\label{tabparspace}List of the variable parameters that
span the parameter space of a two-population IMF studied in this
contribution. Note that $\abd$ is not varied for IC~348 and the
Pleiades since it is not well-constrained by the data for these
clusters.}
\begin{center}
\begin{tabular}{cccc}\hline
Parameter No.&Quantity       &varied from&to (approx.)\\\hline
1            &$\abd$         &0          &1\\
2            &$\mhibd$       &$\mlost$   &$\mlost+0.2$\\
3            &$\log\rpop$&-1         &0\\\hline
\end{tabular}
\end{center}
\end{table}
\begin{table}
\caption{\label{tabdf}Size of the observed sample, $\nsys$
(= the number of observed systems),
number of data points (mass bins), $n_\mathrm{data}$,
number of fitting parameters, $c$, and the number of degrees of
freedom, $\nu=n_\mathrm{data}-1-c$, for the four clusters under study.}
\begin{center}
\begin{tabular}{c|c|ccc}\hline
Cluster      &$\nsys$  &$n_\mathrm{data}$&$c$&$\nu$\\\hline
Trapezium    &1040     &30               &3  &26\\
TA           & 127     &9                &3  &5\\
IC~348       & 194     &9                &2  &6\\
Pleiades     &$\sim500$&19               &2  &16\\\hline
\end{tabular}
\end{center}
\end{table}
\subsection{The Parameter Space}\label{ssec:parspace}
The most straightforward way to calculate the influence of binaries on the
stellar/substellar statistics would be the Monte Carlo method
\citep{KGT91,KTG93}. However,
for better (smoother) results and to reduce the computational efforts we
do not use a Monte Carlo approach here but a semianalytical approach in
which the binary mass function is calculated via numerical integration of
eq. (\ref{defbmf}) for each population. This can be done with a standard
quadrature algorithm. This algorithm has been verified
with a Monte Carlo simulation of a few million random experiments,
each being a random draw from the IMF.
The Monte Carlo method is used later to determine the BD-to-star ratio
and the total binary fractions
within defined mass ranges (\S\S~\ref{ssec:R} and \ref{ssec:f}).
It makes use of the \emph{Mersenne Twister} random number generator
developed by \citet{mersennetwister}. Since there are only a few
runs to be done (in contrast to hundreds of thousands of runs during an
iterated parameter scan, see below) the simple Monte Carlo approach with
an appropriately large random sample is fully sufficient for this purpose.

The lack of star-BD binaries and the truncation of the BD binary separation
distribution suggest two disjunct populations where binary components
are taken only from the same population. For reasons discussed in
\S~\ref{sec:discussion} we call these populations \emph{BD-like} and
\emph{starlike}.
The binary corrections are
therefore applied separately to the stellar and the substellar regime rather
than to a combined population.
Random pairing over the whole mass regime is not considered further as it
leads to too many star-BD binaries \citep{Ketal03}.

The approach here
requires separate application of normalizations on both populations.
For this purpose we define the \emph{population ratio},
\begin{equation}
\rpop=N_1/N_2
\end{equation}
where $N_1$ is the number of individual bodies of the BD-like population
($\mlobd\le m\le\mhibd$) and $N_2$ ($\mlost\le m\le\mmax$)
that of the starlike population.
This must not be mixed up with the \emph{BD-to-star ratio}, $\rbod$,
(eq. [\ref{defR}]) which refers to
\emph{physical BDs} and stars separated by the HBL.

The partial IMF for each population can be
described by parameters $\alpha_i$, $m_j$, and a normalization
constant. In this work a single power law for BDs ($i$ = ``BD'')
and a two-part power law ($i$=1 or 2) for stars is applied.
Thus, there is a mass border, $m_{12}$, separating the two power law
regimes. The parameters of both populations form a three-dimensional
parameter space of the IMF model for each cluster.
It has been found that the lower mass limits of the BD-like population,
$\mlobd=0.01\,\msun$, and that of the starlike population,
$\mlost=0.07\,\msun$, are suitable for all studied clusters.
Furthermore, we focus on the  canonical stellar IMF \citep{Kr01}
with $\alpha_1=1.3$ for $m\le0.5\msun\equiv m_{12}$
and $\alpha_2=2.3$ for $m>0.5\msun$.
The reason is that this canonical IMF has been verified with
high confidence by other observations as well as theoretically.
Thus, only the BD-like power law, $\abd$, the normalization of the
BD-like population against the stellar one, $\rpop$,
and the upper mass border, $\mhibd$, of the BD-like regime are
the parameters to be varied.
Because of the sparse and probably incomplete data sets
for IC~348 and the Pleiades, $\abd$ has also been set to the canonical
value (i.e. $\abd=\alpha_0=0.3$) for these clusters, varying only $\rpop$
and $\mhibd$.

Table \ref{tabparspace} lists the variables and their
range of variation for the Trapezium, IC~348, and the Pleiades.
As the upper mass limits of stars in the clusters one can either take
the maximum observed mass or a theoretical mass limit,
as given by \citet{WK06}.
Because in our model the IMF is cut sharply rather than declining softly
as shown in that work, we here set the mass limit somewhat below the
\citet{WK06} limits.
For the Trapezium we set an upper limit of 10~\tmsun, 1.5~\tmsun\ for TA,
3~\tmsun\ for IC~348 and 5~\tmsun\ for the Pleiades.
Note that the observed most massive star in the Trapezium has a mass of
$\approx50\,\msun$, but varying $\mmax$ to this value does not affect
the results significantly.

Little is known about nonplanetary substellar
objects below the standard opacity limit for fragmentation, 0.01\tmsun\,,
making this a reasonable lower mass limit for our BD statistics.
The lower-mass limit for stars, however, is chosen somewhat
arbitrarily but is justified by a smaller test study that finds less
agreement with observational data used here if the stellar mass is lowered
too much below 0.08\tmsun. As will be shown later, the extreme case
of a stellar mass range that includes the BD mass range leads to
poor fits to the observational MFs
as well as to some inconsistency with the observed binary
fraction. This is discussed later in \S\S~\ref{sec:results} and
\ref{sec:discussion}.
In general, the choice of the mass ranges also influences the
mass ratio distribution,
since for random-pairing each component is distributed
via the IMF. Thus, for a given primary mass out of a chosen population,
the companion mass distribution is just equal to $\xi(m)$ within the mass
interval \mmbox{\mmin\le m\le\mpri}.

The overall binary fraction is taken as a constant for each population.
For the stars we adopt an unresolved binary fraction of 80\pct\
for TA ($\fst=0.80$) and 40\pct\ for the others ($\fst=0.40$),
in accordance to the observational data.
For BDs we choose a value of 15\pct\ for all cases ($\fbd=0.15$). Furthermore,
we found the need to introduce an overlap of the BD and star regimes between
0.07 and about 0.15--0.2\tmsun. Indeed, there is no reason why the
upper mass border of the BD-like population should coincide with the
lower mass border of the starlike population.
The physical implications of this required overlap region are discussed
in \S~\ref{sec:discussion}.

\subsection{$\chi^2$ Minimization}
\label{ssec:chisq}
In \S~\ref{ssec:parspace} a parameter space of three (Trapezium, TA)
and two (IC~348, Pleiades) dimensions of individually adjustable
quantities has been defined.
For each cluster, the quality of a set of fitting parameters
is characterized via the $\chi^2$ criterion,
\begin{equation}
\chi^2=\sum\limits_i^{n_\mathrm{data}} \frac{\left(N_i-N\xis(m_i)\right)^2}{\sigma_i^2}\,,
\end{equation}
where $\xis(m_i)$ is \simf\ at the midpoint of the $i$th bin of the
observational histogram
that is taken as the reference and $n_\mathrm{data}$
is the number of mass bins in the observational histogram
(Tab. \ref{tabdf}).
The values
$\sigma_i$ represent the error bars of the observational data, or,
if none are given, the Poisson errors, and
$N_i$ is the number of systems found in the $i$th mass bin.
The fitted \simf\ is normalized to the total number of systems in the
cluster.

That set of $c$ fitting parameters that minimizes the reduced $\chi^2$
value, $\chi_\nu^2=\chi^2/\nu$, defines our best-fit model. Here
$\nu=n_\mathrm{data}-c-1$ is the number of degrees of freedom.
The values for $n_\mathrm{data}$, $c$, $\nu$, and the sample sizes
for the studied clusters are listed in Table \ref{tabdf}.
The reduction of $\nu$ by 1 is due to the normalization of the
total number of stars and BDs against the observational data.
For the two-component model the fitting parameters are the power in the
BD regime, $\abd$,
the upper mass border of the BD population, and the ratio $\rpop$; thus,
$c=3$ (2, for IC~348 and the Pleiades).
The probability, $P$, that the model has a reduced $\chi^2$,
$\chi_\nu^2=\chi^2/\nu$, as large as or larger than the value actually
obtained is calculated from the incomplete Gamma function \citep{numres},
\begin{equation}
Q(\chi_\nu^2|\nu)=P(\tilde{\chi}_\nu^2\le\chi_\nu^2)\,.
\end{equation}

The logarithmic mass error is not mentioned in the sources, but is given
by the photometric measurements and to a larger degree by uncertain
theoretical models of stars and BDs.
We assumed a value of
\mmbox{\Delta\log m=0.05} for the Trapezium and
\mmbox{\Delta\log m=0.1} for the
others, corresponding to a relative error in the mass estimates of 12\pct\
and 26\pct, respectively. This error has been estimated from the width
of the logarithmic bins in the observational data for the Trapezium, TA and
IC~348. For the Pleiades, for which non-equally spaced data from different
sources are given, we assume the same log mass error as for TA, and
IC~348.
To take this into account the $\log m$ values have
been smoothed by a Gaussian convolution corresponding to a lognormal
smearing of the masses.

After generating the model IMFs for both populations
a binary mass function (eq. [\ref{defbmf}]) is derived separately from the
stellar and
substellar IMF, i.e. there are formally no BD-like companions to starlike
primaries, in accordance with the observations.
This leads to consistency with the observed binary fraction
(\S~\ref{ssec:f}) and the BD desert. Note, though, that as a result of
the required overlap both the BD-like and the starlike population
contain stars as well as BDs; thus, we do have star-BD
pairs in our description (more on this in \S~\ref{sec:discussion}).

The addition of the two
resulting system IMFs leads to an overall \simf\ for the whole mass range.
By adjusting the power law coefficients and the population ratio $\rpop$,
the \simf\ is fitted against the observational data such that $\chi^2$
is minimized. The prominent substellar
peak in the Trapezium cluster below 0.03~\tmsun\ \citep{Muetal02} is ignored
here because it is well below the BD--star mass limit and therefore does not
interfere with any feature there. Furthermore, it is possibly an artefact
of the BD mass-luminosity relation \citep{LaLa03}.

\begin{figure}
\begin{center}
\escl{0.7}
\plotone{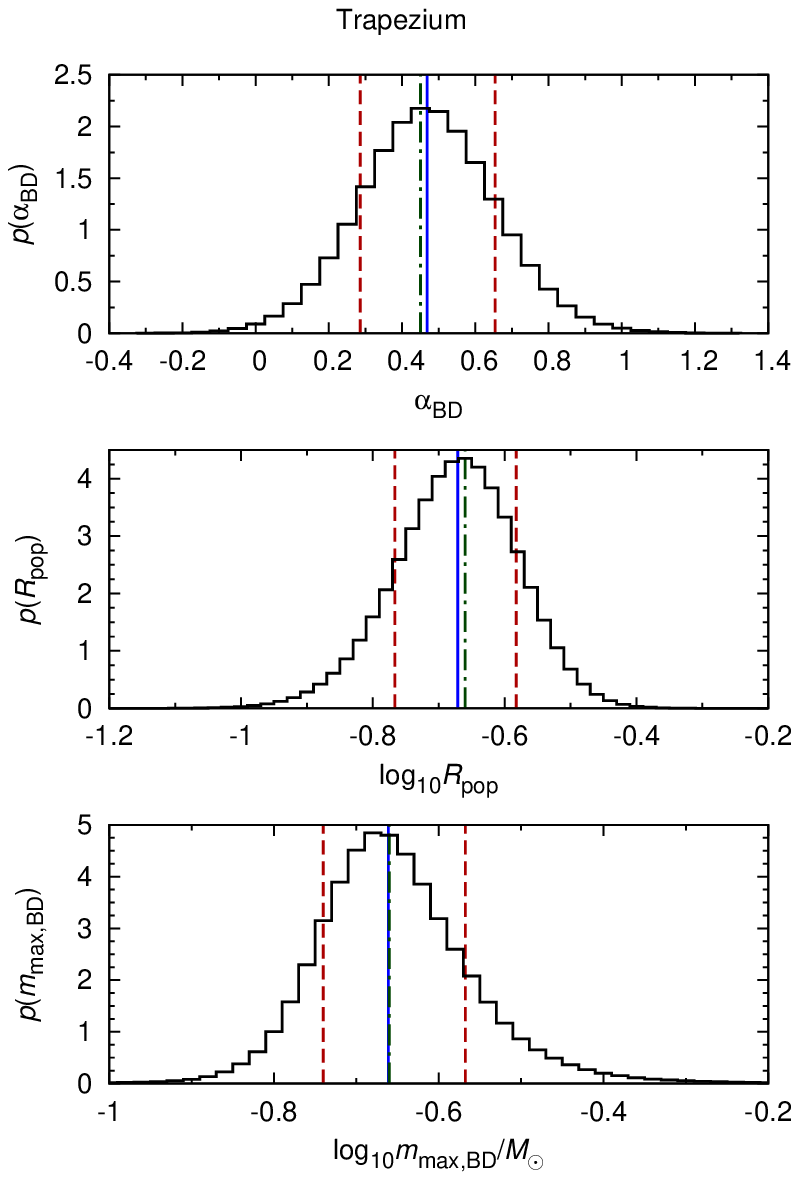}
\caption{\label{pbsstd}%
The marginal probability density distributions of the parameters
$\abd$, $\log\rpop$ and $\log\left(\mhibd/\msun\right)$ as an
estimate of the errors for the Trapezium cluster fit
(see \S~\ref{ssec:errors}).
The peak-width engulfing 68\pct\ of the whole
parameter sample (thus referring to $\approx 1\sigma$,
dashed vertical lines) is taken as the error for each parameter.
Because the median (solid vertical line) of the sample
does not match the best-fit value (dash-dotted line) in all cases,
both are listed in Table \ref{tabalpha}.}
\end{center}
\end{figure}
\subsection{Error Estimation}
\label{ssec:errors}
The errors of the parameters are estimated from the marginal
distribution of each parameter within the parameter space
\citep{STP}. The
marginal probability density distribution, $p$, is
\begin{equation}
\begin{array}{rcl}
p(A_i)&=&\dif P(A_i)/\dif {A_i}\\
      &=&\int\limits_{-\infty}^\infty
         \int\limits_{-\infty}^\infty \mathcal{L}(A_1,A_2,A_3)\,\dif A_k \dif A_j\,,
\end{array}
\end{equation}
where $A_1, A_2$ and $A_3$ correspond to $\abd$, $\rpop$, $\mhibd$
and $i=1,\dots,3$; $j,k\ne i$. The likelihood $\mathcal{L}$
is defined as
\begin{equation}
\mathcal{L}(A_1,A_2,A_3)=\mathcal{N} e^{-\chi^2/2}\,,
\end{equation}
where the normalization constant $\mathcal{N}$ is such that the
integral over the whole range is one.
Note that $\chi^2$ instead of $\chi_\nu^2$ is used here.
The integral is approximated by summation with equidistant
stepping of each parameter.
We conservatively rejected all parameter sets with $P<0.27\pct$
(corresponding to $\chi^2_\nu>\pm3\,\sigma$; \S~\ref{ssec:chisq}).

As an example, the parameter distribution for the Trapezium is
illustrated in Figure \ref{pbsstd}.
The interval around the median which contains 68\pct\ of the
scanned parameter sets is taken as the $\sigma$ measure of the errors.
The interval limits and the medians are shown among the best-fit
results in Table \ref{tabalpha}. Apparently, the median of the
probability density distribution does not always coincide with
the best-fit value. The reason for this is that the distribution
is slightly
asymmetric for most parameters and clusters with the Pleiades
being the by far worst case. This results in asymmetric error bars.
Note that there are sets of $\abd$, $\rpop$, and $\mhibd$ that are
within these error limits but with a $\chi^2>3\sigma$, and thus
the error limits may be slightly over-estimated.

Another possibility to illustrate the statistical significance of
a fit is by its confidence contours within a two-dimensional subspace
of the parameter space, as is shown later in \S~\ref{ssec:R}.

\begin{figure}
\escl{0.5}
\plotone{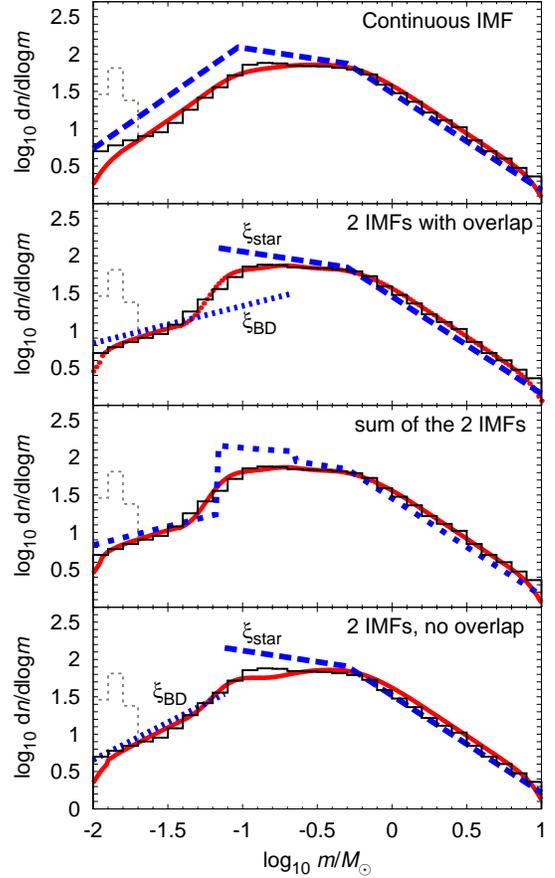}
\caption{\label{IMF4TP}%
Four different IMF models applied to the Trapezium cluster.
Top panel: A classical continuous-IMF fit (dashed line)
for the Trapezium cluster
with random pairing over the whole BD and stellar mass range.
The method of converting the IMF to the \simf\ (solid curve)
is essentially the
same as for the two-component IMFs discussed in this paper but
with no allowance for an overlap or a discontinuity in the BD-star
transition region. Although the general shape of the observational
histogram (thin steps) is represented
by this fit this model leads to an unrealistic high fraction of star-BD
binaries and is therefore discarded.
Second panel: The Trapezium IMF fitted with two separate IMFs,
$\xi_\mathrm{BD}$ (dotted line) and $\xi_\mathrm{star}$ (dashed line),
as in Figure \ref{IMF4sstd} (the two-component IMF).
Third panel: The same but with the sum IMF of both component IMFs
shown here as the possible appearance of the IMF if all binaries could
be resolved. Apparently, there is a ``hump'' between
about 0.07~\tmsun\ and 0.2~\tmsun\ bracketing the
probable overlap region of the BD-like and starlike populations.
Bottom panel: Trapezium fitted with two
separate IMFs but with no overlap. The binary correction leads to
the dip
near the hydrogen-burning limit. However, the fit is only slightly worse
than that one with an overlap.
}
\end{figure}
\begin{figure}
\escl{0.5}
\plotone{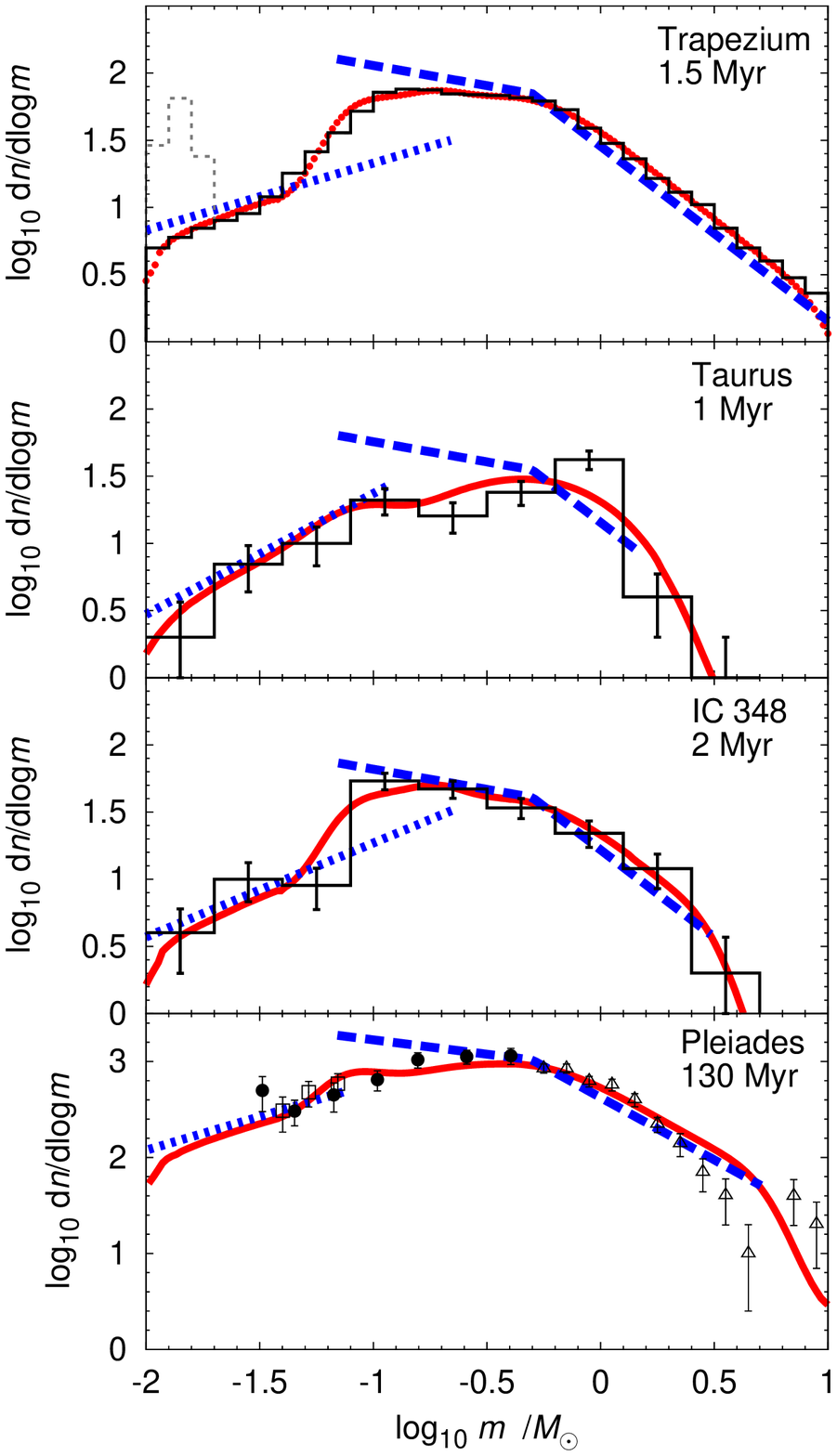}
\caption{\label{IMF4sstd}%
Two-component IMFs of the Trapezium, IC~348, and TA clusters based on
observational data by \citet{Lu04b}, \citet{Luetal03b} and  \citet{Muetal02}
(solid histograms), as well as for the Pleiades cluster using data
from \citet{Moetal03} (filled circles), \citet{Doetal02} (open squares)
and from the Prosser and Stauffer Open Cluster Database (from which only
those for $\log m/\msun\le0.55$ are used for fitting).
For all clusters the stellar IMF, shown here as the dashed line,
has the canonical or standard shape after \citet{Kr01}, i.e. $\alpha_1=1.3$
below 0.5~\tmsun\ and $\alpha_2=2.3$ above 0.5~\tmsun.
The Trapezium IMF is fitted without the
substellar peak indicated by the faint dotted steps.
The dotted lines show the optimized binary-corrected BD-like body
IMFs (distribution of individual BD-like bodies).
The resulting model system IMFs
are represented by the solid curves. Ages are
from \citet{Luetal03a}, \citet{Luetal03b},
\citet{HiCaFe01} and \citet{BSJ04} for TA,
IC~348, Trapezium and the Pleiades, respectively. Note that
the BD-like IMF of IC~348 and the Pleiades are set to the canonical
slope $\abd=0.3$ since they are not well constrained due to
the lack of BD data for these clusters.}
\end{figure}
\begin{figure}
\begin{center}
\escl{1.0}
\epsstr
\plotone{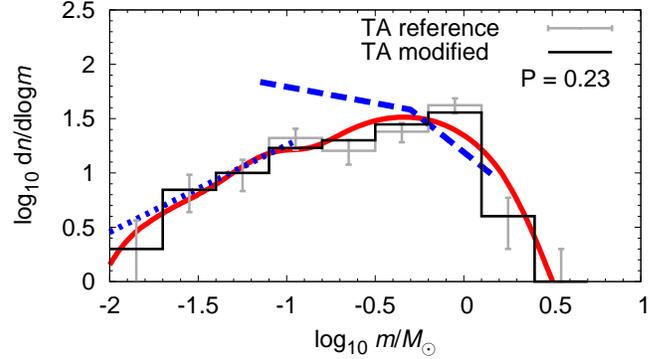}
\caption{\label{TAx}%
The two-component model with canonical stellar IMF for a slightly
modified TA observational histogram (black solid steps). The peak at 1~\tmsun\
and the dent around 0.3~\tmsun\ have been slightly flattened without
leaving the error bars of the original data (grey steps and bars).
This fit has a reasonable confidence of about $P=0.25$, about 12 times
larger than for the original histogram ($P=0.02$, see Tab. \ref{tabalpha}).
}
\end{center}
\end{figure}
\begin{figure}
\escl{0.5}
\plotone{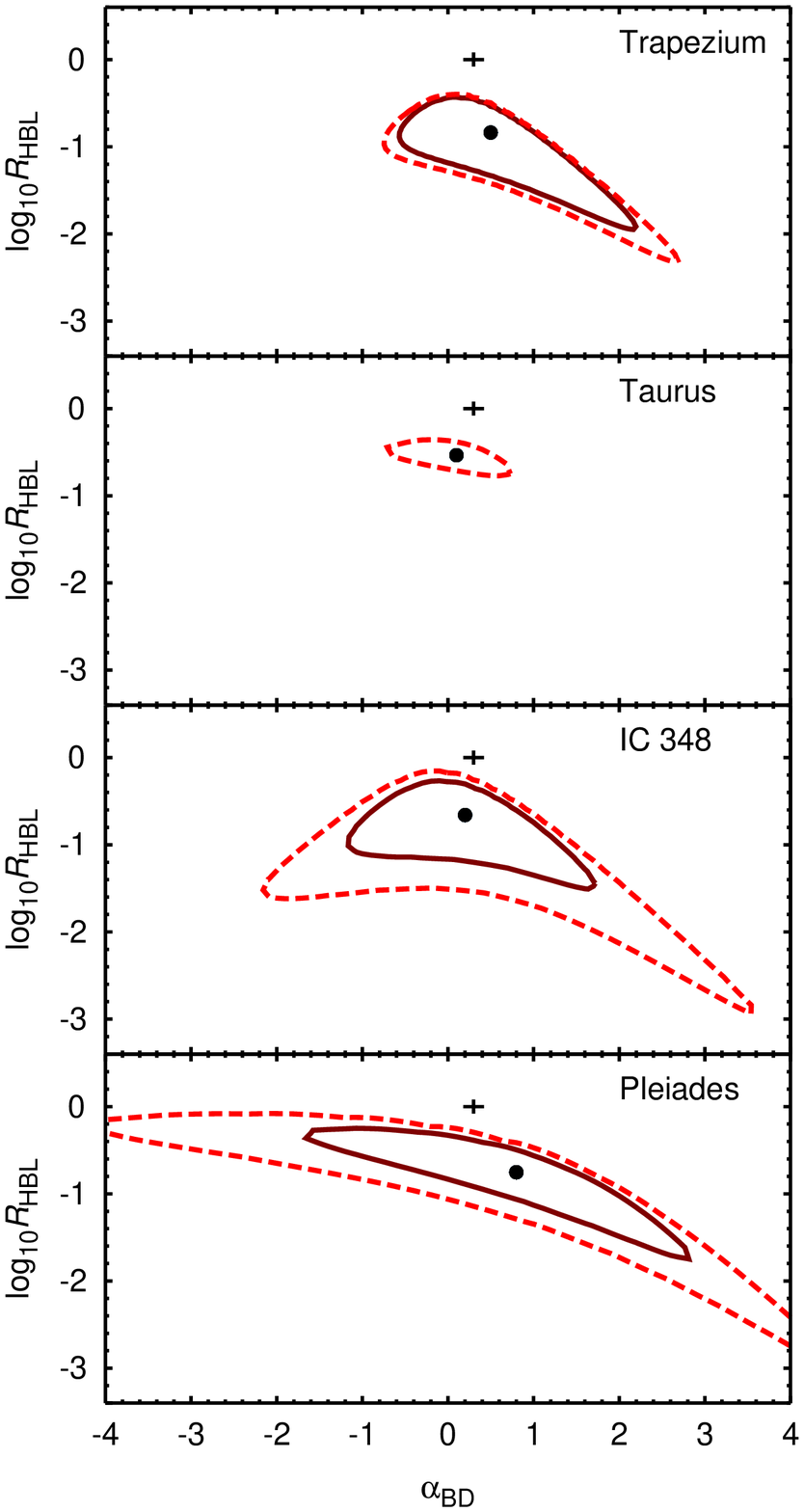}
\caption{\label{cnt4}%
Contour plots of 5\pct\ and 1\pct\ significance levels from $\chi^2$
fitting (see \S~\ref{sec:method} for details) of
$\abd$ and the BD-like to starlike ratio
$\rkbd$ at $m=0.075\,\msun$ (eq. \ref{defrkbd})
for the Trapezium, TA, IC~348 and the Pleiades.
The stellar IMF is canonical, i.e. $\alpha_1=1.3$ and $\alpha_2=2.3$
while the upper BD mass limits, $\mhibd$, are the best-fit values
from Table \ref{tabalpha}.
All fitting parameters outside the
solid line are rejected with 95\pct, and those outside the
dashed line are rejected with 99\pct\ confidence. The optimum
of the Trapezium and TA is marked by the filled circle while
the cross marks the standard/canonical configuration with a continuous
IMF (i.e. $\log\rkbd=0$) and $\abd=0.3$.
As can be seen, it is well outside both levels
(even for arbitrary $\abd$). For the Pleiades it is still well outside
at least for reasonable choices of $\abd$.
Note that the optimum for $\abd$ for IC~348 and the Pleiades does exist
in this 2D cross section but not in the full 3D parameter space.}
\end{figure}
\begin{figure}
\begin{center}
\escl{0.50}
\plotone{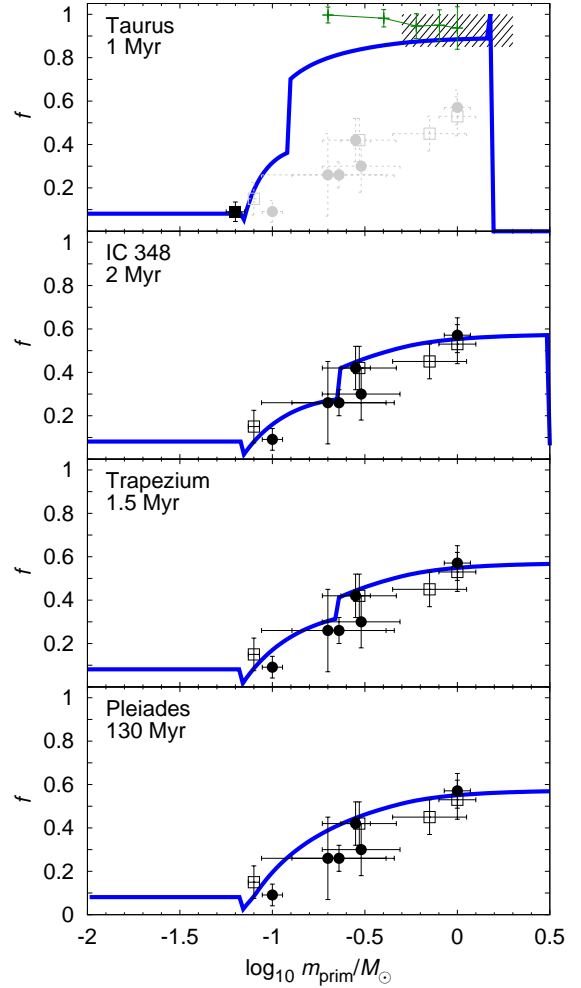}
\caption{\label{fbin4}%
The binary fraction $f$ (eq. \ref{fmpri})
of the four discussed clusters as a function of
primary mass, $m_\mathrm{prim}$, for the two-component best-fit
IMF model (with canonical stellar IMF) is shown by the solid line.
The BD binary fraction appears flat due to the equal-mass pairing used in
our algorithm. Since merely very sparse data is available for low-mass BDs,
only the higher-mass end of the curve should be taken into account.
For comparison,
the average binary fractions of G, K, M and L type stars in the solar
neighbourhood
(average age $\sim 5$~Gyr; filled circles: \citealt{Lada2006}, open squares:
\citealt{Ketal03}) have been added to each panel.
The shaded region in the top panel indicates the approx. 1~Myr old
pre-main sequence data \citep{Duchene99} while the thin solid line
represents the initial TA-like dynamical model from \citet{Ketal03}.
The single filled square in the top panel at $\log m\approx-1.2$
is the VLMS/BD datum inferred by \citet{Krausetal06} for TA.
Note that the stars in TA have a binary fraction near
100\pct\ \citep{Duchene99,Ketal03}.}
\end{center}
\end{figure}

\section{Results}
\label{sec:results}
\begin{table*}
\caption{\label{tabalpha}The best-fit BD-like power law coefficients, $\abd$,
population ratios, $\rpop$, and BD-like upper mass limits, $\mhibd$.
The uncertainties are derived in \S~\ref{sec:method}.
Note that $\abd$ is set to the canonical value for IC~348 and the Pleiades.}
\begin{center}
\begin{tabular}{|l|ccc|ccc|cc|}\hline
           &\multicolumn{3}{c|}{Median $\pm1\sigma$}&\multicolumn{3}{c|}{Best fit}  &&\\
Cluster    &$\abd$                 &$\log\rpop$       &$\mhibd$               &$\abd$            &$\log\rpop$&$\mhibd$ &$\chi_\nu^2$&$P$\\\hline
Trapezium  &$0.47\pma{0.18}{0.19} $&$-0.67\pma{0.10}{0.09}$&$0.22\pma{0.04}{0.05}$ & 0.45             &-0.66           &0.22     &0.23&0.99998\\      
TA         &$-0.08\pma{0.63}{1.05}$&$-0.68\pma{0.13}{0.14}$&$0.11\pma{0.03}{0.04}$ & 0.00             &-0.62           &0.12     &2.69&0.020   \\      
IC~348     &---                    &$-0.57\pma{0.23}{0.19}$&$0.22\pma{0.06}{0.08}$ &$0.3\,\mbox{(c)}$ &-0.51           &0.22     &1.01&0.416   \\      
Pleiades   &---                    &$-0.82\pma{0.44}{0.11}$&$0.07\pma{0.02}{0.20}$ &$0.3\,\mbox{(c)}$ &-0.80           &0.06     &1.20&0.257   \\\hline
\end{tabular}
\end{center}
\end{table*}

\subsection{The IMF for BDs and Stars}
\label{ssec:IIMF}
Figure \ref{IMF4TP} demonstrates the results for different models
(continuous IMF, two-component IMF with and without overlap region) for the
Trapezium cluster. The continuous
IMF is shown for illustration only. Its underlying model assumes
a single population containing BDs as well as stars \citep{Ketal03}.
Instead of an overlap
as for the two-component model the mass border between the regime of
the BD slope $\abd$ and the (canonical) stellar regime is varied. The
total binary fraction, $\ftot$, is set to 0.4 with random pairing
among the entire population. Our calculations give
a best fit with $\abd=-0.4\pm{0.2}$, $\mhibd\equiv\mlost=0.093\pm{0.01}$
(and canonical stellar IMF) which is comparable to the canonical
IMF for BDs and stars \citep{Kr01}.
As already mentioned in \S~\ref{sec:intro}, it leads to a high number
of star-BD binaries, $\fmix=15\pct\pm4\pct$ altogether (see \S
\ref{ssec:f}),
while \citet{Ketal03} found 7\pct--15\pct\ star-BD binaries with
$a\ga30\unit{AU}$ in their standard model with stars and BDs.
Furthermore, the different orbital
properties of BD-BD and star-star binaries (Fig. \ref{sepdist})
cannot be explained by a single-population model without further
assumptions.
Therefore, it is not considered further here.

The two-component IMF model (Fig. \ref{IMF4TP}, \emph{second panel})
accounts for the empirical binary properties of BDs and stars.
Apart from this it also fits the observational data
(Fig. \ref{IMF4TP}, \emph{histogram}) slightly better than
the continuous model, especially in the BD/VLMS region and the
``plateau'' between 0.1 and 0.5~\tmsun.

Objects with equal mass and composition appear as equal in observations,
even if they have been formed in different populations. Thus, it is hard
to determine their formation history.
A high-resolution survey that
resolves most of the multiples would yield an overall mass function
from the lowest mass BDs to the highest mass stars.
However, if such a resolved individual mass function is composed of two
overlapping populations, it may be possible to detect its imprints
as an excess of objects within the overlap region.
To illustrate this ((Fig. \ref{IMF4TP}, \emph{third panel}),
we construct such an overall IMF by simply adding the two
IMF components from the second panel. BDs and stars are paired separately
from their respective mass range.
It is clear that the addition of the two IMFs leads to discontinuities,
i.e. in the overlap region the continuous IMF increases steeply at the
minimum stellar mass and drops again at the upper mass limit of the
BD-like regime.
Even with a smoother drop at each end of the BD-like and starlike IMF
it is likely to be detectable once most of the multiples have
been resolved. Thus, a discovery of such a ``hump'' within the
resolved IMF would strongly support the two-component model.

To illustrate the effect of the overlap region, a two-component IMF
without an overlap is shown in the bottom panel of Figure \ref{IMF4TP}.
Like the continuous
IMF this approach gives a slightly worse fit, especially between
0.1 and 0.2~\tmsun\ ($-1\le\log m/\msun\le-0.7$). However,
the ``dip'' in this region is only weak.

The results of the two-component IMF model
for all clusters are shown in Figure \ref{IMF4sstd}, the power law
coefficients being listed in Table \ref{tabalpha}.
For the Pleiades and IC~348 the BD IMF slope has been kept
standard, i.e. $\abd=0.3$ \citep{Kr01},
because the sparse, and in the case of the Pleiades
most probably incomplete data, do not allow useful confidence
limits to be placed on $\abd$. For the same reason
the highly uncertain observational data for $\log(m/\msun)\ge0.65$ have
not been used for fitting (but are still plotted for completeness).

As can be seen the BD/VLMS and stellar IMFs do not meet at the BD-star
boundary. The number density of individual BDs near the BD-star border is
about one third of that of individual low-mass M dwarfs. This discontinuity
cannot be seen directly in the observational data because it is masked by
the different binary fractions for different masses (\S
\ref{ssec:IMF:SIMF}), but the IC~348 data
(histogram in the third panel of Figure \ref{IMF4sstd})
may indeed be showing the discontinuity (compare third panel of Figure
\ref{IMF4TP}).

Furthermore, there are large uncertainties
in the mass determination of stars in stellar groups such as TA which
may probably lead to even larger error bars of the bins than shown in the
observational data. Even within the given error bars certain variations of
these data provide a much better fit to the canonical stellar IMF
with $P=0.25$ (original: $P=0.02$, Table \ref{tabalpha}). This
has been done in Figure \ref{TAx} by lowering the peak near 1~\tmsun\
and somewhat rising the ``valley'' around 0.3~\tmsun
corresponding to the re-shuffling of 9 stars out of 127
(Tab. \ref{tabparspace})
by 0.1 to 0.4~\tmsun\
(one bin width), e.g. through measurement errors.
This suggests that, in agreement with \citet{Ketal03}, the TA IMF might
not necessarily be inconsistent with the canonical IMF.

Except
for the Pleiades all clusters show features near the peak in the low-mass
star region that are slightly better fitted with a separate BD population
\emph{and} an overlap region. Although these features alone do not reject
the continuous IMF model, they might be taken as a further support
of the argumentation towards a two-populations model.

It should be mentioned that the continuous IMF model (Fig. \ref{IMF4TP},
\emph{first panel})
still fits the observed Trapezium and Pleiades MF with high confidence
(while failing for the other clusters) but leads to VLMS binary
properties that are inconsistent with the observed properties as
mentioned in \S~\ref{sec:intro}. A possible extension of our modelling
would thus be to fit both the observed IMF \emph{and} the observed binary
statistics.

\subsection{BD to Star Ratio}
\label{ssec:R}
We analyzed the BD-to-star ratio $\rbod$ of TA, Trapezium and IC~348,
defined as
\begin{equation}\label{defR}
\rbod=\frac{N(\mbox{0.02--0.075}\,\msun)}{N(\mbox{0.15--1.0}\,\msun)}\,,
\end{equation}
where $N$ is the number of bodies in the respective mass range.
The mass ranges are chosen to match those used in \citet{KB03b} in their
definition of $\rbod$. Since
BDs below 0.02~\tmsun\ are very difficult to observe and since TA and
IC~348 do not host many stars above 1~\tmsun\ (in contrast to Trapezium and the
Pleiades), we restricted the mass ranges to these limits.

The Pleiades cluster is difficult to handle due to a lack of BD data.
Moreover,
at an age of about 130~Myr \citep*{BSJ04} it has undergone dynamical evolution,
and massive stars have already evolved from the main sequence which
affects the higher mass end of the IMF \citep{Moetal04}.

\begin{figure}
\begin{center}
\escl{1.0}
\epsstr
\plotone{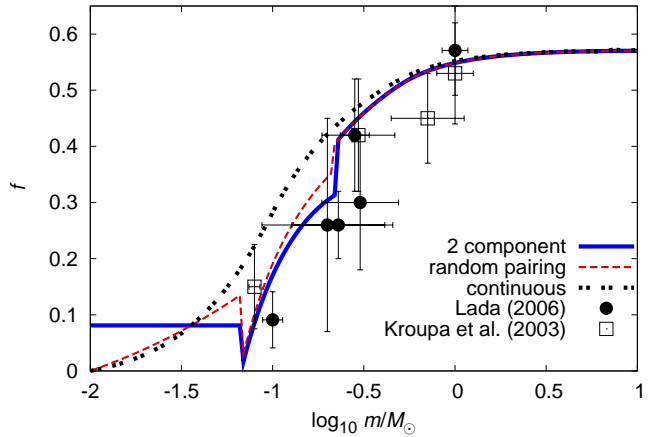}
\caption{\label{fbuni}%
The binary fraction $f(\mpri)$ of our best 2-component IMF fit for the
Trapezium (solid line) with $\abd=0.5$ in
comparison with the binary fraction, $f_\mathrm{c}(\mpri)$, of the best
continuous IMF model (dotted line, see also top panel in Figure \ref{IMF4TP}).
The 2-component model assumes random pairing of stars but
equal-mass pairing for BD binaries. The continuous model, on the other hand,
assumes random pairing of companions chosen randomly from the canonical IMF
($\abd=0.3$, $\alpha_1=1.3$ and $\alpha_2=2.3$).
This model is merely shown for illustrative purposes and has already been
discarded (\S~\ref{sec:intro}). The thin dashed line refers to the same
two-component IMF as the solid line, but with random-pairing instead of
equal-mass pairing for BDs.
The visible jumps in $f(\mpri)$ at
$\log(m/\msun)=-1.15$ and -0.7 are due to the truncation
at the upper and lower mass limit of the BD-like and the starlike
population with their different binary properties.
The global shapes of the continuous and the two-component IMF curves
are similar but $f_\mathrm{c}(\mpri)$ is significantly higher
than $f(\mpri)$ in the low-mass star and VLMS regime and outside the
uncertainties of the data by \citet{Lada2006}.}
\end{center}
\end{figure}

Another relevant quantity is the ratio $\rkbd$ of BD-like to starlike
objects at the HBL, i.e. the classical BD-star border,
\begin{equation}\label{defrkbd}
\rkbd=\frac{\xi_\mathrm{BD}(0.075\,\msun)}{\xi_\mathrm{star}(0.075\,\msun)}\,.
\end{equation}
In the classical continuous IMF approach it is one by definition, because
otherwise the IMF would not be continuous.
In a two-component IMF its value depends on the shapes of the
BD-like and the starlike IMF, as well as on the BD-to-star ratio.
The evaluation of equation (\ref{defrkbd}) yields
$\rkbd=0.17$ for the Trapezium, $\rkbd=0.30$
for TA, $\rkbd=0.22$ for IC~348, and $\rkbd=0.3$ for the Pleiades.
Although $\rkbd$ is not an input parameter here but is calculated from
the two IMFs and their relative normalization, $\rpop$, it
could easily be used as one instead of $\rpop$ while calculating the
latter from $\rkbd$.

Figure \ref{cnt4} shows the contour plots of the 5\pct\ and 1\pct\
significance ranges in the $\abd-\rkbd$ space for
Trapezium, TA, IC~348, and the Pleiades.
The significance values are calculated from $\chi_\nu^2$ via the incomplete
Gamma function as described in \S~\ref{sec:method}.
The contours mark the regions outside which the
hypothesis of a two-component IMF with a single power law for BDs
and a double power law for stars has to be rejected with 95\pct\
or 99\pct\ confidence.
Also shown (by a cross) is the standard configuration with
$\abd=0.3$ and $\log\rkbd=0$ for the continuous standard IMF.
This point is outside both levels for all three clusters, and at least
for Trapezium and TA it is well outside even for arbitrary
$\abd$.
In other words, \emph{the corresponding hypothesis
of a continuous IMF has to be rejected with at least 99\pct\ confidence.}

The size of the non-rejection areas can be used for an estimate of the
errors of $\abd$ and $\rkbd$. However, one has to keep in
mind that these are only maximum possible deviations with all the other
parameters kept at the optimum. The non-rejection areas in the
full three-dimensional parameter space are therefore
expected to be somewhat narrower.

\subsection{Binary Fraction}
\label{ssec:f}
Additionally, the binary fraction for each cluster as well as the
total binary fraction is calculated for the best-fit models.
The fraction of binaries as a
function of the primary mass, \mmbox{f(m_\mathrm{prim})}, among all systems
is shown in Figure \ref{fbin4}.
For stars the binary fraction is a monotonic function of the
primary mass in agreement with the data \citep{Lada2006}, at least
for the Trapezium, IC~348, and the Pleiades.
For BDs, however,
it is flat due to the equal-mass pairing. In the case
of random pairing it would approach zero for very low mass BDs
(Fig. \ref{fbuni}).
The true
mass ratio function grows monotonically with the mass ratio and becomes very
steep near $q=1$, as shown by \citet{Reidetal06}. Thus, for BDs the true binary
fraction is probably closer to the equal-mass case than the random pairing
case.

For comparison, Figure \ref{fbuni} also shows the binary fraction,
$f_\mathrm{c}(\mpri)$, for the Trapezium that would result from a
continuous IMF. Although the overall shape is very similar to that of the
two-component model the binary fraction near the BD-star transition is
significantly higher for low-mass stars than the observed values while
being approximately equal for stars above 1~\tmsun.
Thus, a continuous IMF cannot fit the observational data as good as a
two-population IMF even if $\ftot$ is reduced.

The total fractions of BD-BD binaries, $\fbd$, of
star-star binaries, $\fst$, and the fraction of
(very low mass) star-BD binaries, $\fmix$, are of further interest.
Pairs of the latter type consist of
two objects of the BD-\emph{like} or star\emph{like} population (see
\S~\ref{sec:discussion} for the motivation and definition of the
populations) but where the primary object is a star
($m_\mathrm{prim}\ge0.075\,\msun$)
within the BD-VLMS overlap region between 0.07 and 0.15~\tmsun,
while the companion is a true, physical BD with a mass below
0.075~\tmsun.

For each cluster we define
\begin{eqnarray}
\fbd &=&\frac{\nbbd}{\nsbd}\\
\fmd &=&\frac{\nbmd}{\nsmd}\\
\fst &=&\frac{\nbst}{\nsst}\\
\fmix&=&\frac{\nbsb}{\nsst}\,,
\end{eqnarray}
where $\nbbd$ is the number of all BD-BD binaries (i.e. all objects
have masses $\le0.075\,\msun$), $\nbmd$ that of all M dwarf--M dwarf
(MD-MD) binaries, $\nbst$ that of all
star-star binaries, and $\nbsb$ the number of mixed (star-BD) binaries
(composed of a VLMS and a BD).
Furthermore, $\nsbd$ is the
number of all BD systems (including single BDs), $\nsmd$ that of all M dwarf
systems, and $\nsst$
the number of all systems with a star as primary and with a stellar, BD,
or no companion.
As for the BD-to-star ratio we also
applied the (primary) mass ranges from \citet{KB03b}
on the BD and star sample from each cluster but with no gap between the
BD and the stellar regime,
i.e. $0.02\,\msun\le\mpri\le 0.075\,\msun$ for BDs
and $0.075\,\msun\le\mpri\le 1\,\msun$ for stars. In addition, the binary
fraction of M dwarfs is calculated in the same way as the stellar one
but in the mass range $0.075\,\msun\le\mpri\le 0.5\,\msun$.
The uncertainty limits of the binary fractions and BD-to-star ratios
have been derived from those of the IMF slopes by applying the
minimum and maximum slopes.

The results are shown in Table \ref{tabbinfrac}.
The binary fractions for BDs vary only
slightly between 12\pct\ and 15\pct, while the stellar
binary fraction, $\fst$, is about 70\pct\ for TA and 30\pct--40\pct\
for the others. They are apparently slightly lower than the binary
fractions that are set for the starlike population because of the
nonconstant distribution binary as shown in Figure \ref{fbin4}.
The binary mass function (eq. [\ref{defbmf}]) is smaller in the
mass region below 1~\tmsun\ and thus the binarity is
below average if the focus is set on this region.

Furthermore, in the overlap region the relatively low binary fraction of
VLMSs from the BD-like population also contributes to $\fst$,
which results in an even lower value of $\fst$. This trend is more
emphasized for the M dwarf binary fraction, $\fmd$, which is about 10\pct\
lower than $\fst$ for each cluster but still much larger than $\fbd$.

The star-BD binary fraction $\fmix$ is of special interest since it is
the measure for the ``dryness'' of the BD desert.
Note that due to the equal-mass pairing used here for BD-like binaries, the
BD-like population formally does not contribute to $\fmix$.
All star-BD binaries are from the starlike regime which
extends down to $\mlost=0.07\,\msun$, i.e. into the BD mass regime.
For the Trapezium, IC~348, and the Pleiades our two-component models
yield values between 2\pct\ and 2.5\pct, whereas TA shows
$\fmix\approx5\pct$. For comparison, the continuous IMF from the top panel
of Figure \ref{IMF4TP} corresponds to $\fmix=15\pct\pm4\pct$.
For mass ratio distributions other than equal-mass pairing
we expect a higher $\fmix$, consisting mostly of binaries from the BD-like
population with primary masses slightly above the HBL and companion
masses below.
For random pairing of BD-like binaries this increment is between
0.01 and 0.03. Also the lower mass limit of the starlike population
slightly influences $\ftot$. 
It should also be noted that the size of the error limits for $\fmix$
are calculated from the uncertainties of the IMF model
(Tab. \ref{tabalpha}) but do not include the uncertainties of $\mlost$.

\begin{table*}\small
\caption{\label{tabbinfrac}
Binary fractions $\fbd$ for BD-BD, $\fmd$ for MD-MD,
$\fmix$ for star-BD and $\fst$ for star-star
binaries in the studied clusters. Note that the star-BD binaries
result from the pairing in the overlap region
between about 0.075 and 0.15~\tmsun.
Thus, for example, in IC~348 2.1\pct\ of all stars are expected
to have a physical BD as a companion. In addition, the BD-to-star
ratio between 0.02 and 1.0~\tmsun\, $\rbod$, and at the
HBL, $\rkbd$, is listed. Note that the numerical uncertainties of
$\fmix$ are probably much smaller than true ones since $\fmix$ largely
depends on $\mlost$.
}
\begin{center}
\begin{tabular}{|l@{\quad}|@{\quad}c@{\quad}c@{\quad}c@{\quad}c@{\quad}|c@{\quad}c@{\quad}|}\hline
Cluster  &$\fbd       $&$\fmd$        &$\fst$        &$\fmix$       &$\rbod$       &$\rkbd$\\\hline
Trapezium&$0.13\pm0.01$&$0.30\pm0.01$&$0.34\pm0.01$&$0.023\pm0.002$&$0.18\pm0.03$&$0.17\pm0.04$\\
TA       &$0.15\pm0.01$&$0.64\pm0.06$&$0.69\pm0.05$&$0.046\pm0.005$&$0.27\pm0.09$&$0.30\pm0.11$\\
IC~348   &$0.13\pm0.02$&$0.29\pm0.02$&$0.33\pm0.02$&$0.021\pm0.004$&$0.20\pm0.10$&$0.22\pm0.11$\\
Pleiades &$0.13\pm0.02$&$0.33\pm0.03$&$0.37\pm0.03$&$0.025\pm0.002$&$0.28\pm0.18$&$0.28\pm0.18$\\\hline
\end{tabular}
\end{center}
\end{table*}

\section{Discussion: Brown dwarfs as a separate population?}
\label{sec:discussion}
\subsection{An Apparent Discontinuity}
By correcting the observed MFs for unresolved multiple systems
a discontinuity in the IMF near the BD/VLMS region emerges.
We have also tried to model continuous
single-body IMFs, but we find this hypothesis of continuity to be
inconsistent with the observed MFs given the observational data on the binary
properties of stars and BDs.
We have
shown that the \emph{empirically determined} difference in the binary
properties between BDs/VLMSs on the one hand side and stars on the other,
and the \emph{empirical finding} that stars and BDs rarely pair, implies a
discontinuity in the IMF near the BD/VLMSs mass.

Thus, the discontinuity
in binary properties, which has already been interpreted to mean two
separate populations \citep{KB03b}, also implies a discontinuity in the
IMF. This relates the probably different formation mechanism more
clearly to the observational evidence.

We have performed a parameter survey allowing the IMF
parameters $\abd$, $\rpop$, and $\mhibd$ to vary finding that the canonical
stellar IMF ($\alpha_1=1.3$ and $\alpha_2=2.3$) cannot be discarded even for
TA, and that the BD/VLMS discontinuity is required for all solutions.
The discontinuity uncovered in this way, if measured at the classical
BD-star border, is of a similar magnitude
for the stellar clusters studied ($0.17\le\rkbd\le0.30$),
supporting the concept of a universal IMF, which is, by itself,
rather notable.

We recommend calling these populations \emph{BD-like} and \emph{starlike}
with respect to their formation history. These populations have probably
overlapping mass ranges since there is no physical reason for the upper
mass limit of the BD-like population to match the lower mass limit
of the starlike one. Furthermore, the best-fit models suggest such an
overlap between 0.07~\tmsun\ and about 0.2~\tmsun.
According to this classification through the formation history,
BD-like objects would include VLMSs,
while starlike ones would include massive BDs.

The overlap region
implies that BD-like pairs can consist
of a VLMS-BD pair, and that a starlike binary can consist of a
stellar primary with a massive BD as a companion.
As can be seen from Table \ref{tabbinfrac}, the star-BD fractions
are similar (except for the dynamically unevolved TA) and
that the star-BD binary fraction, $\fmix$, is about 2\pct--3\pct.
This is somewhat
higher than the value of $<1\pct$ BD companion fraction \citet{GreLin06}
found for nearby stars but still far below the value expected for a single
population model (at least 7\pct\--15\pct\ in \citealt{Ketal03} and
$15\pct\pm4\pct$ in our calculations). Simulations by \citet{BBB03},
\citet{BaBo05}, and \citet{Bate2005}
predict a fraction of about 2\pct\ star-BD binaries
(actually one M dwarf with a BD companion out of 58 stars formed in three
independent calculations).
Note that, in our model, $\fmix$ is a prediction of
the required overlap region and is sensitive to the overlapping range.
As our model for the Pleiades IMF suggests, the overlapping range might
be considerably smaller than that we have found for the other clusters.
Furthermore, we did not fit the lower mass border, $\mlost$, of the
starlike population but simply assumed a value of 0.07~\tmsun\ which
is well in the BD mass regime (and is the major source of star-BD binaries
in our Pleiades model). A slight increase of $\mlost$ by only
0.01~\tmsun\ would cause the Pleiades star-BD binary fraction to drop
to nearly zero.

Several authors doubt the existence of two separate populations.
Most recently,
\citet{EiSt07} summarize that the observational community in general
prefers the model of starlike formation for BDs.
They mention the detection of isolated
proto(sub)stellar ``blobs'' in the Ophiuchus B and D clouds, which
may support the theory of starlike formation for BDs.
It remains unclear, though, how many
of these blobs will actually form BDs instead of dissolving from lack
of gravitational binding energy.
\citet{GoWi07} refer to a private communication with {\AA}. Nordlund
stating that the pure turbulence theory predicts about 20,000
transient cores for every actual pre-stellar core of about 0.1~\tmsun.

We recall that one of the main reasons for the existence of a separate
population is the semi-major axis distribution of BD binaries (Fig.
\ref{sepdist}). But \citet{Luetal07} also mention a wide binary
BD in Ophiuchus with a separation of approximately 300~AU.
Indeed, a small number of wide BD binaries are known.
However, it can be doubted that the occasional discovery
of a wide BD binary may expand the narrow semi-major axis distribution
(Fig. \ref{sepdist}) to a starlike one. The striking evidence posed
by the lack of BD companions to stars is a strong indication for two
populations. It is usually ignored by the community, though.
We note that even if there actually may be some BDs that formed
starlike they are most probably a minority.

There is also the interpretation of the BD desert being a
``low-$q$ desert'' rather than an absolute mass-dependent drop in the
companion mass function. \citet{GreLin06} find a low-mass companion
desert of solar-type primary stars between approximately 0.01 and
0.06~\tmsun. They find this interval to be dependent on the primary
star mass and therefore predict M dwarfs to have BD companions,
and that M dwarfs ought to have a
companion desert between a few Jupiter masses and the low-mass BD
regime. However, this interpretation does not address the different
orbital properties of BDs and stars as well as the different
$q$ distribution.

\subsection{Implications for the Formation History}
Can the existence of such a discontinuity, i.e. the formation
of two separate populations, be understood theoretically?
Although BDs and stars appear to be distinct populations the formation
of BDs is likely to be connected to star formation.
Bate et al. (2002, 2003) show that BDs
significantly below 0.07~\tmsun\ cannot form in a classical
way since the minimum mass they need for stellar-type formation would
also lead to progressive accretion and growth to stellar mass unless they
are in regions with very low mass infall rates.
But such regions are very rare, because the prevailing densities
and temperatures cannot achieve the required Jeans masses,
as also stressed by \citet{GoWi07}.
For this reason, \citet{AdFa96} expected BDs to be rare.
Only a small fraction of the BDs, especially those at the high-mass end
of the BD regime, may form this way if the surrounding gas has been
consumed by star formation processes just after the proto-BD has reached
the Jeans mass.
To explain the actually higher BD frequency (per star) in recent
surveys the accretion process has to be terminated or
impeded somehow \citep*{PPVBoetal}.

Also the above-mentioned
differences in the distribution of the semimajor axes between BDs and stars
cannot be explained by a scaled-down star formation process, because
that would imply a continuous variation and a much broader semi-major axis
distribution for BDs and VLMSs that has not been observed \citep{Ketal03}.
While binary stars show a very broad distribution of their semimajor axes
peaking at about 30~AU the semimajor axes of BDs are
distributed around about 5~AU with a sharp truncation at 10--15~AU
(Fig. \ref{sepdist}). No smooth transition region between both regimes
can be recognized \citep{Cloetal03}. In a high angular resolution
survey \citet{LaHoMa07} found that the orbital radius distribution
of binaries with V-K $<$ 6.5 appears to differ significantly
from that of cooler (and thus lower mass) objects, suggesting a sudden
change of the number of binaries wider than 10~AU at about the M5 spectral
type. This is in agreement with our finding of a possible BD-like population
that extends beyond the hydrogen-burning mass limit into the VLMS
regime.

In a radial-velocity survey of Chamaeleon I, \citet{Joergens2006a} has
found evidence for a rather low binary fraction below 0.1~AU, while most
companions found in that survey orbit their primaries within a few AU.
For this reason an extreme excess of close BD binaries that cannot be
resolved by imaging surveys appears to be unlikely.
A larger binary fraction than about 15\pct\ would thus not be plausible.
\citet{BasRei2006} suggest an upper limit of $26\pct\pm10\pct$
for the BD binary fraction based on their own results
($11\pct^{+0.07}_{-0.04}$, for separations below 6~AU) and the survey
by Close et al. (2003; $15\pct\pm7\pct$, for separations greater than 2.6~AU)
by simple addition of the results. This is nearly consistent with
a BD binary fraction of 15\pct, since the survey is neither magnitude-
nor volume-limited. However, they admit that their value may be
over-estimated since
the objects with separations between 2.6 and 6~AU are counted twice.
We note further that even a total BD binary fraction of 25\pct\
($\fbd=0.25$), although outside the error limits of our best-fit models,
would only lead to a minor change in the fitted IMFs.

It has been argued \citep{BasRei2006} that the lower binary fraction of BDs
is just the extension of a natural trend from G dwarfs to M dwarfs
(Figs. \ref{fbin4} and \ref{fbuni}).
Our contribution has shown that this trend can be understood by the
simple fact that there are many fewer possibilities to form a binary near
the lower mass end than for higher component masses.
The observational data are in better agreement with a minimum mass, $\mmin$,
near the hydrogen-burning mass limit and a low overall binary fraction
of BD-like objects than with an ``all-in-one'' IMF from the lowest mass BDs
to the upper stellar mass limit, as shown in Figure \ref{fbuni}. This
observed trend thus appears as an additional enforcement of the
two-populations model of BDs and stars.

Given that the conditions for a starlike formation of BDs are very rare
\citep{BBB03}, four alternative formation scenarios for BDs apart
from starlike formation can be identified, namely
\begin{enumerate}
      \item Formation of wide star-BD binaries via fragmentation of a
            proto- or circumstellar disk and subsequent disruption
            by moderately close encounters.
      \item Formation of BDs as unfinished stellar embryos ejected from their
            birth system.
      \item Removal of the accretion envelopes from low-mass protostars via
            photoevaporation.
      \item Removal of the accretion envelopes due to extremely close stellar
            encounters \citep{PriPod95}.
\end{enumerate}
Scenario 4 can be ruled out as the major BD formation mechanism because the
probability of such close encounters,
with required flyby distances typically below 10~AU \citep{KB03b}
for efficient disruption of accretion envelopes (less than a tenth of
those proposed by Thies et al. [2005] for triggered planet formation),
is far too low for such a scenario being a significant
contribution to BD formation.
The photo evaporation model, as studied by \citet{WiZi04},
also cannot be the major mechanism of BD formation \citep{KB03b}. It
predicts a variation of the IMF with the population number and
density of the host
cluster. In dense starburst clusters (young globular clusters)
with a larger number of
O/B stars or even modest clusters such as the ONC with a dozen O/B stars
compared
to TA, a larger fraction of low-mass stars would have halted in
growth. This would
result in a bias towards M dwarfs, since many of them would be
failed K or G dwarfs.
In contrast to this prediction, \citet{Brietal02} and \citet{Ketal03}
show that the IMFs of TA
and ONC are very similar in the mass range 0.1--1~\tmsun,
while globular clusters likewise have a low-mass MF similar
to the standard form \citep{Kr01}.

\subsection{Embryo Ejection}
\label{ssec:ejection}
\citet{Rei00} and \citet{ReiCla01} introduced the formation of BDs as
ejected stellar embryos as the alternative scenario 2.
If a forming protostar in a newborn
multiple system is ejected due to dynamical instability
its accretion process is terminated
and the object remains in a protostellar state with only a fraction of
the mass compared to a fully developed star. Since the final mass
is physically independent of the hydrogen fusion mass limit
one would not expect the mass range of ejected
embryos to be truncated at the HBL and thus expect an
\emph{overlap region} between these populations.
This fully agrees with the requirement
of having to introduce such an overlap region in order to fit the
observed \simf\ in \S~\ref{sec:method}.

This model gives some hints to understand the low BD binary fraction as
well as the truncation of the semimajor axis distribution of BDs.
The decay of a young multiple system
of three or more stellar embryos typically leads to the ejection of
single objects but also to the ejection of a small fraction of close
binaries. In order to survive the ejection, the semi-major axis of
such a binary must be significantly smaller (by a factor of about 3)
than the typical orbital separation within the original multiple system.
A similar explanation is that the orbital velocity of the
BD binary components has to be higher than the typical ejection
velocity in order to keep the interaction cross section of the binary with
other system members small. Indeed, the velocity dispersion of BDs in the
embryo-ejection model shown in \citet{KB03b} is
$\lesssim 2\unit{km\,s^{-1}}$ for the majority of the BDs. This is in
good agreement with the Keplerian
orbital velocity of each BD-binary member of about
1.5--$2\unit{km\,s^{-1}}$ for an equal-mass binary of 0.05--0.08~\tmsun\
and $a\approx 10\unit{AU}$. The majority of BD binaries have smaller
separations, and, consequently, higher orbital velocities and are bound
tighter. This would set the low binding energy cut near
$E_\mathrm{bind}=0.2\unit{pc^{-2}\,Myr^2}$ in Figure \ref{edist}.

There have been numerical simulations of star formation and dynamics,
e.g. by \citet{BBB03} and \citet{Umetal05}, in which binaries are
produced via ejection that show remarkably similar properties to the
actually observed ones. \citet{Umetal05} describe the formation of
BDs from decaying triple systems. Their simulations predict a semi-major
axis distribution between about 0.2 and 8~AU (see Fig. 8 in their paper),
peaking at 3~AU.
This is slightly shifted towards closer separations compared with
the results by \citet{Cloetal03} but still in agreement with the
observational data. In contrast to this, \citet{GoWi07} doubt
the frequent formation of close BD binaries via ejection, arguing
that hydrogen-burning stars which formed via ejection were almost
always single.

In further qualitative support of the embryo ejection
model, \citet{Guetal06} describe a deficit by a factor of 2
of BDs near the
highest density regions of TA relative to the BD abundance in
the less dense regions that can
possibly be explained by dynamical ejection and consequently larger
velocity dispersion of stellar embryos, i.e. BDs.
A starlike fragmentation scenario would
result in an opposite trend since the Jeans mass is smaller for higher
densities, thus allowing gas clumps of lower mass to form (sub)stellar
bodies. Contrary to this, \citet{Lu06} did not find any evidence for a
different spatial distribution of BDs and stars in TA.
Recently, \citet{KumSch07} have found that substellar objects in both
the Trapezium and IC~348 are distributed homogeneously within twice the
cluster core radii while the stellar populations display a clustered
distribution.
They conclude that these distributions are best explained with a higher
initial velocity dispersion of BDs, in accordance with \citet{KB03b},
supporting the embryo-ejection model.

However, the embryo-ejection model has a challenge to reproduce the
high fraction of significant disks around BDs that have been observed in
young clusters. Several studies, e.g. \citet{Nattaetal04},
reveal a considerable number of BDs with an infrared
excess that indicates the presence of warm circum(sub)stellar material.
While these studies do not show the actual mass of these disks because
a small amount of dust in these disks is sufficient to produce these
excesses, \citet*{SJW06} found 25\pct\ of BDs in TA having disks with
radii $>10$~AU and significant masses (larger than 0.4~$M_\mathrm{J}$).
For the remaining 75\pct\ no disks were detectable.
In simulations, e.g. by \cite{BBB03}, such large disks
survive occasionally, but less frequently (about 5\pct)
than suggested by the observations \citep{SJW06}.
However, \citet{SJW06} point out that their results to not rule out the
embryo-ejection scenario and admit that this mechanism may still be relevant
for some BDs. In general, the embryo-ejection model is in agreement with
at least the existence of low-mass circumsubstellar disks up to about 10~AU
\citep{BBB03,Umetal05,PPVGuetal}. \citet{TKT05} also
show that a thin low-mass disk can survive near-parabolic prograde
coplanar encounters above about
three Hill radii with respect to the disk-hosting BD.
This means that a disk with a radius up to about 5~AU (later
viscously evolved to about 10~AU, see \citealt{BBB03}) can survive
a flyby of an equal-mass embryo within about 15~AU while larger
disks or widely separated binaries would be disrupted.

Furthermore, \citet{BBB03} suggest from their
simulations that the binary fraction via ejection might by as small
as about 5\pct\ (see also \citealt{WhietalPPV}).

\subsection{Disk Fragmentation and Binary Disruption}
\label{ssec:diskfrag}
The fragmentation of protobinary disks with subsequent disruption
of a star-BD binary is another promising alternative scenario.
A disk can fragment during the accretion process if it reaches a
critical mass above which the disk becomes gravitationally unstable
against small perturbations. Fragmentation may
be triggered by an external perturbation, i.e. infalling gas clumps
or a passing neighbor star. The latter mechanism may also be capable of
triggering fragmentation in relatively low-mass circumstellar disks
resulting in rapid planet formation \citep{TKT05}.
Thus, BD formation via disk fragmentation
likely plays an important role in the early ages of the cluster where the
frequency of massive disks is highest \citep{HLL2001, WhietalPPV}.

Following the argumentation of \citet{GoWi07}, disk fragmentation may
explain the observed distribution of BD binary separations at least as
well as the embryo-ejection model. In addition, it may explain the
existence of wide star--BD binaries, since a fraction of the initial
wide binaries can survive without being disrupted. Because the likelihood
of disrupting close encounters depends on the mass and density of the
host cluster, one expects a higher fraction of those wide star-BD
binaries in smaller and less dense clusters and associations like TA.
However, further studies and observations are needed
to test this hypothesis.

\citet{WhiSta06} show that BDs can actually form as widely
separated companions to low-mass stars with mass $m$ at a sufficiently
large disk radius, $r_\mathrm{disk}$,
\begin{equation}
r_\mathrm{disk} \gtrsim 150\unit{AU}\left(\frac{m}{\msun}\right)\,,
\end{equation}
where the disk is cool enough to allow a substellar clump to undergo
gravitational collapse. For a primary star below 0.2~\tmsun\ this
minimum radius therefore becomes less than 90~AU which is in remarkable
agreement with the two wider VLMS binaries found by \citet{Konetal07}.
Such large distances to the primary star allow the formation of BD-BD
binaries as well as the survival of circumsubstellar disks
up to about 10--30~AU, depending on the total mass of the
presubstellar core and the mass of the primary star.
Furthermore, this scenario explains the existence of wide star-BD binaries.
In addition, such a wide star-BD binary can be disrupted by moderately
close encounters of about 100--200~AU (i.e. a distance similar to the
star-BD orbital radius), the disruption of such systems appears to be
likely in contrast to the disruption of accretion envelopes
as required in the already rejected scenario 4.

\subsection{Summary}
Both the embryo-ejection model and disk fragmentation with
subsequent wide binary disruption explain the above-mentioned
connection between stars and BDs, since BDs start to form like stars before
their growth is terminated due to their separation from their host system
or from lack of surrounding material in the outer parts of a circumstellar
disk. It is obvious that the formation
rate of these embryos is proportional to the total star formation rate.

For these reasons these formation mechanisms appear to be the most likely
ones for BDs and some VLMSs and BD/VLMS binaries. It cannot be decided
yet which scenario is the dominant mechanism. This may depend on the
size and the density of the star-forming region.
We expect, on the other hand, the classical starlike formation
scenario to be of some importance only for the most massive BDs.

The possibility of two different \emph{alternative} BD formation
mechanisms (disk fragmentation and embryo ejection) may
lead to another discontinuity in the intermediate mass BD IMF
since both scenarios correspond to different binary fractions
as mentioned above. The currently available data, however, are far from
being sufficient for a verification of this prediction.

\section{Conclusions}
\label{sec:conclusions}
The different empirical binary properties of BDs and stars strongly
imply the existence of two separate but mutually related populations.
We have shown that if the IMF of BDs and stars is analyzed
under consideration of their binary properties then
there is a discontinuity in the transition
region between the substellar and stellar regime that is quite
independent of the host cluster.
The discontinuity in the IMF near the HBL
is a strong logical implication of the disjunct binary properties and
suggests splitting up the IMF into two components, the BD-like and the
starlike regime. An alternative but equivalent description would be to
view the stellar IMF as a continuous distribution function ranging from
about 0.07 to 150~\tmsun\ \citep{WK04}, and a causally connected but
disjoint distribution of (probably mostly) separated ultra--low-mass
companions and ejected embryos with masses ranging from
0.01~\tmsun\ to 0.1--0.2~\tmsun.
While the canonical stellar IMF is consistent with
the observed stellar MFs at least for the Trapezium, IC~348, and the
Pleiades, the sub-stellar IMF of at least the Trapezium and TA has a
power law index that is consistent with the canonical value $\abd=0.3$.
Within the error limits, our analysis does not reject the canonical
power law indices for BDs and stars for any cluster.

The discontinuity is often masked in the observational data due to a
mass overlap of both populations in the BD-VLMS region
as well as the higher apparent masses of unresolved binaries compared to
single objects in the observed \simf.
The discontinuity in the number density near the
HBL is a step of approximately a factor of 3--5
(Table \ref{tabbinfrac}). This implies
a general dependency between both populations and is, as far as we
can tell, consistent with the
scenario of disrupted wide binaries \citep{GoWi07} as well as with the
truncated-star scenario (e.g. as an ejected stellar embryo,
\citealt{ReiCla01}), since the
number of unready stars is directly correlated with the total amount
of star formation in the host cluster.
Both embryo-ejection and wide binary disruption are also consistent
with the properties of close binary BDs.

Our results (Tab. \ref{tabalpha})
suggest that about one BD is produced per 4--6 formed stars.
This suggests the necessity
of a distinct description of BDs and stars as well as the connection between
these two populations through their formation process.

\section*{Acknowledgements}
This work was partially  supported by the AIfA and partially by DFG grant
KR1635/12-1.

\end{document}